\documentclass[manuscript]{aastex}

\usepackage{amsmath,amssymb}

\slugcomment{submitted to the Astrophysical Journal}

\shorttitle{CME self-similar expansion and Lorentz self-forces}
\shortauthors{Subramanian et al.}

\begin{document}


\title{Self-similar expansion of solar coronal mass ejections: implications for Lorentz self-force driving}


\author{Prasad Subramanian\altaffilmark{1,2} and K. P. Arunbabu}
\affil{Indian Institute of Science Education and Research, Dr Homi Bhabha Road, Pashan, Pune - 411008, India}

\author{Angelos Vourlidas}
\affil{Space Science Division, Naval Research Laboratory, 4555 Overlook Ave SW, Washington, DC 20375, USA}

\and

\author{Adwiteey Mauriya}
\affil{Indian Institute of Science Education and Research, Dr Homi Bhabha Road, Pashan, Pune - 411008, India}


\altaffiltext{1}{p.subramanian@iiserpune.ac.in}
\altaffiltext{2}{Center for Excellence in Space Sciences, India (http://www.cessi.in/)}


\begin{abstract}
We examine the propagation of several CMEs with well-observed flux rope signatures in the field of view of the SECCHI coronagraphs aboard the STEREO satellites using the GCS fitting method of Thernisien, Vourlidas \& Howard (2009). We find that the manner in which they propagate is approximately self-similar; i.e., the ratio ($\kappa$) of the flux rope minor radius to its major radius remains approximately constant with time. We use this observation of self-similarity to draw conclusions regarding the local pitch angle ($\gamma$) of the flux rope magnetic field and the misalignment angle ($\chi$) between the current density ${\mathbf J}$ and the magnetic field ${\mathbf B}$. Our results suggest that the magnetic field and current configurations inside flux ropes deviate substantially from a force-free state in typical coronagraph fields of view, validating the idea of CMEs being driven by Lorentz self-forces.
\end{abstract}





\section{Introduction}
Since Earth-directed coronal mass ejections (CMEs) from the Sun are typically held responsible for most major geomagnetic storms, a thorough understanding of the forces governing their initiation and propagation are of considerable practical importance. CME kinematics exhibit a variety of characteristics \citep{yashiro04, webb12}. They range from very fast CMEs that experience most of their acceleration within $\approx$ 1--2 $R_{\odot}$ above the solar limb to ones that show evidence of being continuously driven throughout typical coronagraph fields of view that extend upto $\approx$ 30 $R_{\odot}$ \citep{sv07}. CMEs (especially the ones whose mechanical energies are increasing through the coronagraph fields of view) are commonly thought to be driven by Lorentz self-forces (e.g., \citealp{sng13, olmetal, chkr, linetal98, chn96, kmrst96}). We interpret these ${\mathbf J} \times {\mathbf B}$ forces as being due to misaligned currents and magnetic fields contained within the evolving flux rope structure. On the other hand, the ``drag'' forces contributing towards CME deceleration are thought to be due to momentum coupling between the CMEs and the ambient solar wind (e.g., \citealp{gpl00, lew02, cgl04, vrs, slb12}). While we have recently arrived at a preliminary understanding of the physics underlying the drag force (\citealp{slb12}), we do not yet have a very good understanding of the details of the driving force. In fact, we do not even have a clear idea of the typical heliocentric distance at which the driving force ceases to be important (in comparison to the drag force).

The magnetic energy contained by CMEs is generally thought to be responsible for propelling them; this concept was quantitatively demonstrated by \citet{vr00}. \citet{sv07} showed that, on the average, the magnetic energy contained in CMEs provides at least 74 \% of what is required for their propagation from the Sun to the Earth. We identify a set of well observed CMEs observed by the SECCHI coronagraphs  \citep{howard08} aboard the STEREO satellites \citep{kaiser08}. A large majority of CMEs observed with coronagraphs are now confirmed to possess a flux rope morphology (e.g., \citealp{angelos13, jie13}). Accordingly, we fit the graduated cylindrical model for flux rope CMEs (\citealp{thr09}) to these well observed CMEs in order to obtain their 3D structure. As we will discuss below, one overarching result of this fitting procedure is that the flux rope CMEs evolve in a manner such that the ratio of their minor to major radii remains constant. Although there has been some observational evidence for the fact that some flux rope CMEs expand in a self-similar manner in the coronagraph field of view  (e.g., \citealt{poom10, kilpua2012, robin2013}), our work is the first systematic demonstration of self-similar expansion. 
Self-similarity has been invoked in a number of theories relating to CME propagation (e.g., \citealp{kmrst96, Dem, wang09, olmedo2010}). 
Using the observed self-similar expansion, we draw conclusions regarding the extent to which the flux rope structures are non-force-free. In \S~2 we describe the observational results, which we use to draw conclusions regarding the current and magnetic field configurations in \S~3. Conclusions are drawn in \S~4. 

\section{Data analysis}

\begin{figure}
\includegraphics[width = \columnwidth]{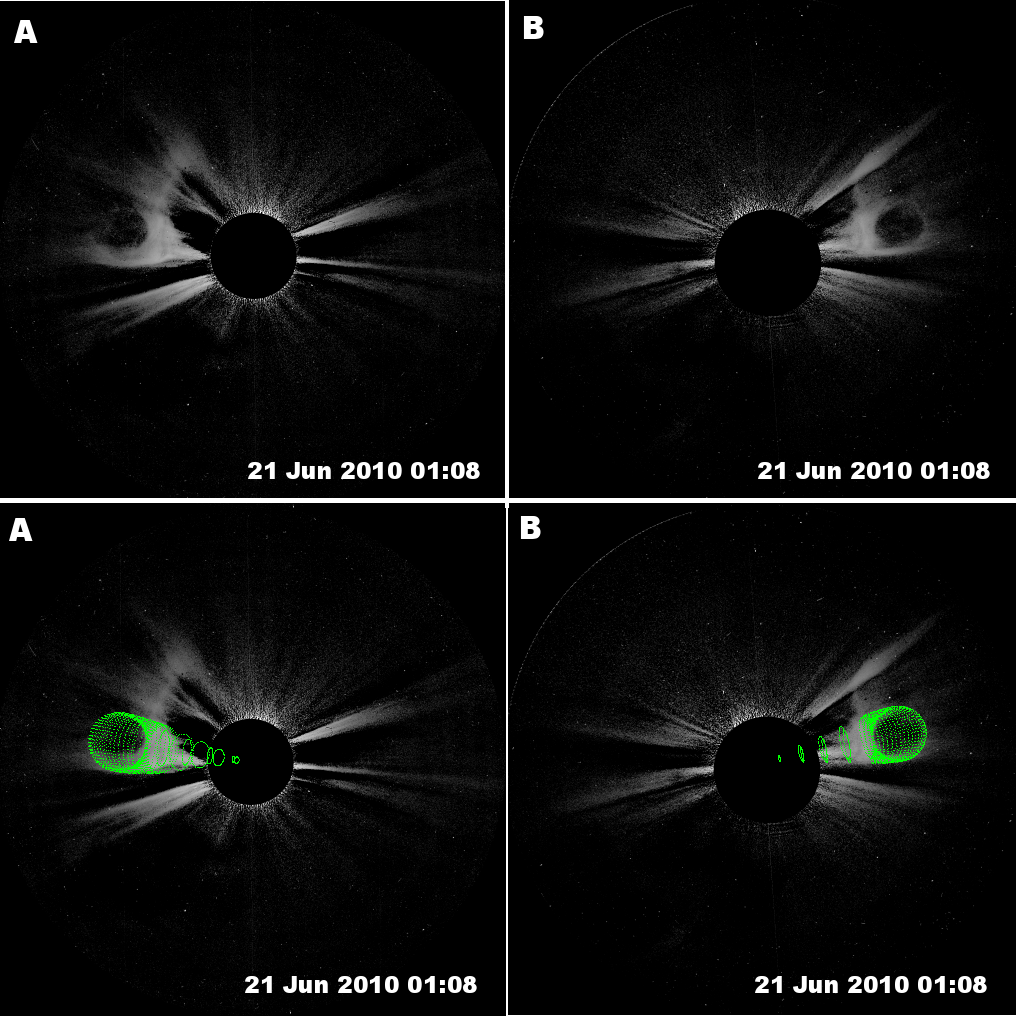}
\caption{A screenshot of a CME on 21 June 2010 that illustrates the flux rope fitting procedure. The left panels are from STEREO A and the right panels are from STEREO B. The upper panels show the white light CME data and the lower panels have the flux rope (displayed as a green wiremesh) structure superposed.}
\label{crd}
\end{figure}

We have identified 9 well observed CMEs with the SECCHI A and B coronagraphs aboard the STEREO spacecraft. Following the method outlined in \citet{thr09} we fit a three-dimensional geometrical flux rope configuration to the images in SECCHI A and B coronagraphs simultaneously at each timestamp. We show a representative screenshot in Figure~\ref{crd}, which shows the flux rope fitting for the images at 01:08 UT on 21 June 2012. At each timestamp, we have taken care to ensure that cross section of the flux rope is fitted only to the dark cavity visible in the coronagraph images.  The flux rope fitting routine yields a variety of geometrical parameters. Table \ref{tbl1} summarizes the most relevant ones for each of the events we have studied in this paper. We have followed CMEs only as far as it is possible to make a clear, unambiguous fit to the flux rope model. This means that we can follow some CMEs farther out than others, which is why some CMEs have many more timestamps than others. For the purposes of this paper, the main results from Table \ref{tbl1} are:

\begin{itemize}
\item 
The quantity $\kappa$, which is the ratio of the flux rope minor radius to its major radius, remains approximately constant with time for a given CME. This conclusion holds for all the CMEs we have studied, and is a clear demonstration of the fact that flux rope CMEs expand in a nearly self-similar manner.

\item
 The values of $\kappa$ for different CMEs in Table \ref{tbl1} are: $0.44 \gtrsim \kappa \gtrsim 0.2$ 
\end{itemize}

For a given CME, it may be noted that the geometrical flux rope fitting procedure we use allows for different values of $\kappa$ at different timestamps. There is no assumption of self-similarity built into this procedure, and the approximate self-similarity observed in CME evolution is thus physical. Several of the flux rope CMEs studied by \citet{kilpua2012} using the GCS method (that we use here) also evolve in a self-similar manner, with $0.39 \gtrsim \kappa \gtrsim 0.23$. Even in the HI field of view, similar measurements of CMEs for the events used by \citet{robin2013} also reveal self-similar expansion, with $0.60 \gtrsim \kappa \gtrsim 0.25$. \citet{sv09} found that the subset of CMEs from \citet{sv07} that were subject to a net driving force show a constant value of $\kappa$; they used this fact to derive the axial current enclosed by these flux rope CMEs. We note that \citet{sv09} used LASCO data, which did not have the advantage of two viewpoints that the current study does. They selected CMEs that seemed to propagate mostly in the plane of the sky, and interpreted the circular cross-section visible in LASCO images as the cross-section of the flux rope. While we only study flux rope CMEs that expand in a self-similar manner here, we note that there are CMEs whose expansion is not self-similar (e.g., \citealp{cheng2014}). This might be because their evolution is genuinely non self-similar, or an illusion arising out of CME rotation (e.g., \citealp{angelos2011}).
We next turn our attention to the implications of the observed self-similar propagation in the context of a flux rope model where the evolution is governed entirely by Lorentz self-forces.

\section{Lorentz self-forces in flux ropes}
\subsection{Self-similar expansion}
 A qualitative sketch of a fluxrope configuration is shown in figure \ref{flrp}. The subject of Lorentz self-forces in flux ropes has a long history, starting from \citet{sha66} through treatments like \citet{anz, garchen, chn96, kmrst96, chkr, sv09, olmetal}  to mention a few. Broadly, they all appeal to variants of ${\mathbf J} \times {\mathbf B}$ forces, arising from the currents and magnetic fields carried by the flux rope structure. The assumption of self-similar flux rope evolution is built into several popular theoretical treatments of Lorentz self-force driving (e.g., \citealp{kmrst96, sv09, olmedo2010}). In the treatment of \citet{kmrst96}, self-similar evolution is a consequence of assuming that axial magnetic flux and helicity are both conserved. However, they do not use a specific value of the self-similarity parameter \begin{equation}
\kappa \equiv \frac{a}{R}
\label{eqkappa}
\end{equation} in their treatment.

We next examine some other treatments involving Lorentz self-forces  (e.g., \citealp{chn96, chkr}) that do not explicitly appeal to self-similar expansion. Since its often difficult to specify unique magnetic field and current configurations for a non force-free flux rope structure (see \citealp{chn12} for some examples), several authors have used the self-inductance of a slender, axisymmetric, circular flux rope as a starting point. This quantity (in cgs units) is \citep{sha66,lanlif}

\begin{equation}
L = 4\, \pi\, R \biggl [ {\rm ln}\biggl (\frac{8\,R}{a}\biggr ) - 1 \biggr ]\, ,
\label{eqinductance}
\end{equation}
where $R$ is the major radius of the flux rope and $a$ is its minor radius. The magnetic energy associated with a current loop such as this carrying an axial current $I$ is $U_{m} = (1/2) L I^{2}$. The Lorentz self-force acting along the major radius is then derived as

\begin{equation}
f_{R} = \frac{1}{c^{2}}\, \frac{\partial}{\partial R} U_{m} = \frac{2\, \pi\,I^{2}}{c^{2}}\,\biggl [ {\rm ln}\biggl (\frac{8\,R}{a}\biggr ) - 1 \biggr ]\, ,
\label{eqselfforce1}
\end{equation}
where $c$ is the speed of light. Thereafter, the force per unit arc length acting along the major radial direction is calculated as $f_{\rm L} = (1/2 \pi R) f_{R}$. It may be noted that the last step in equation~(\ref{eqselfforce1}) can be arrived at only if the quantity $\kappa \equiv a/R$ is assumed to be constant. In other words, any treatment that uses equation~(\ref{eqselfforce1}) implicitly assumes that the flux rope evolves in a self-similar manner.  However, we note that some treatments (e.g., \citealp{chn96, chkr}) use equation~(\ref{eqselfforce1}) (and therefore implicitly assume self-similar expansion) and yet have separate differential equations for the evolution of the flux rope major radius ($R$) and its minor radius ($a$).

Here, we use the observed values of the self-similarity parameter $\kappa$ to determine the relation between the local pitch angle of the magnetic field configuration inside the flux rope and the misalignment angle between the current density and the magnetic field.

\subsection{How misaligned are ${\mathbf J}$ and ${\mathbf B}$?}

\begin{figure}
\includegraphics[width = \columnwidth]{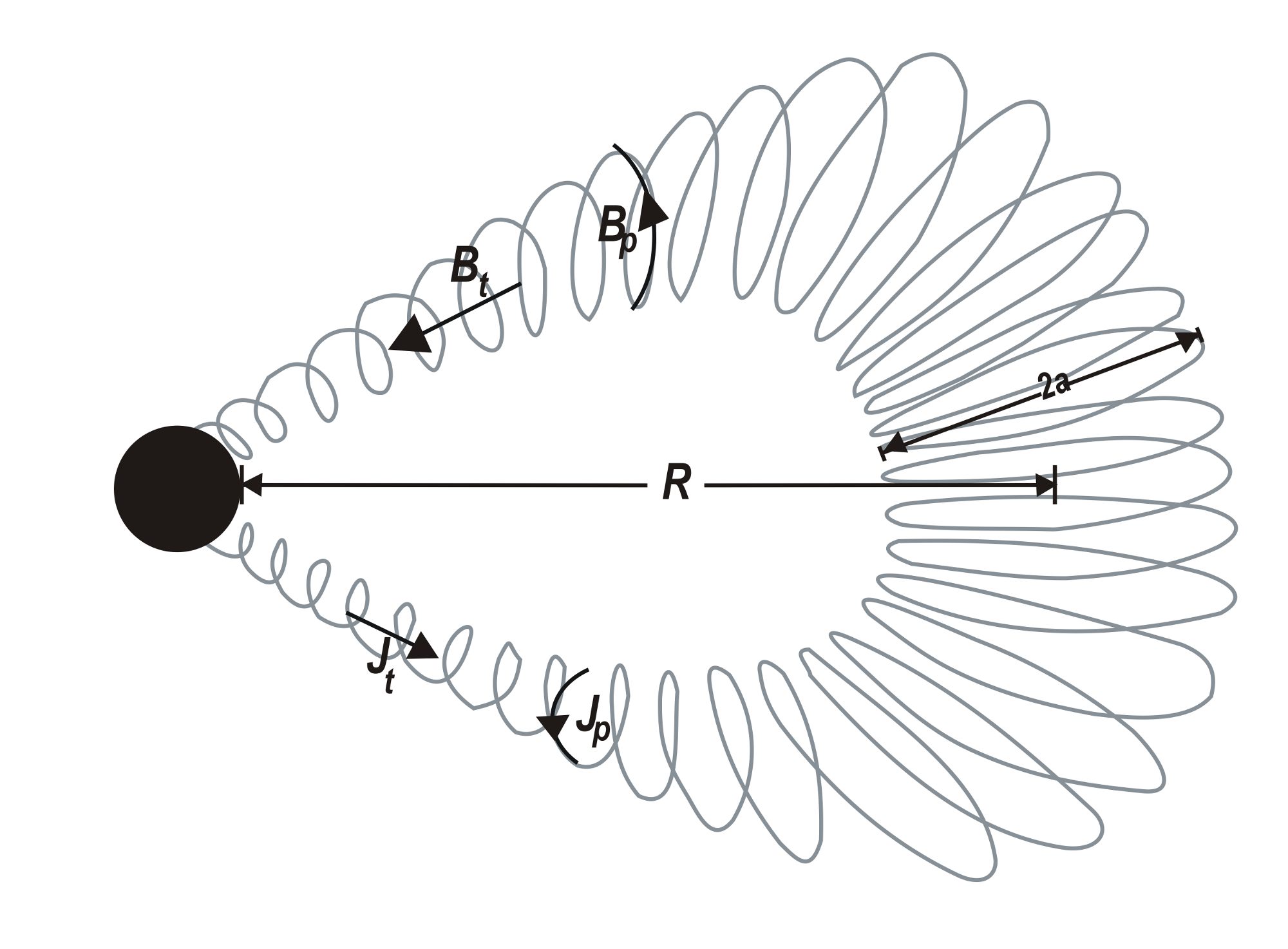}
\caption{A schematic of the fluxrope magnetic field. The fluxrope minor radius is $a$ and its major radius is $R$. The directions of the toroidal and poloidal current densities and magnetic fields are indicated.}
\label{flrp}
\end{figure}

Since Lorentz self-force driving necessarily involves non-force free configurations, our first step is to evaluate the angle between the current density ${\mathbf J}$ and the magnetic field ${\mathbf B}$. We decompose the current and magnetic field into poloidal and toroidal components:

{\begin{eqnarray}
\nonumber
 \mathbf{J} = J_p \mathbf{i_p} + J_t \mathbf{i_t} \equiv c_1 B_t \mathbf{i_p} + c_2 B_p \mathbf{i_t} \\
 \mathbf{B} = B_p \mathbf{i_p} + B_t \mathbf{i_t} \, ,
\label{eqdecompose}
\end{eqnarray}}
where ${\mathbf i_p}$ and ${\mathbf i_t}$ are unit vectors in the poloidal and toroidal directions respectively (see fig \ref{flrp}).
Defining
\begin{equation}
c_{1} \equiv J_{p}/B_{t}\,\,\,\, {\rm and} \,\,\,\, c_{2} \equiv J_{t}/B_{p}, 
\label{eqc1c2def}
\end{equation}
and the magnetic field pitch angle $\gamma$

\begin{equation}
\gamma \equiv \tan^{-1} \frac{B_p}{B_t}\, .
\label{eqpitchangle}
\end{equation}

the angle $\chi$ between the current density ${\mathbf J}$ and the magnetic field ${\mathbf B}$ can be written as

{\begin{eqnarray}
\nonumber
\sin \chi = \frac{| \mathbf{J} \times \mathbf{B}|}{|\mathbf{J} ||\mathbf{B}|} = \\
\frac{1 -  \frac{c_{2}}{c_{1}}\, \tan ^2 \gamma}{\biggl [(1 + \tan ^2 \gamma) \, (1+ \frac{c_2^2}{c_1^2} \tan ^2 \gamma) \biggr ]^{1/2}} \, ,
\label{eqsinchi}
\end{eqnarray}}
where we have used Equation (\ref{eqdecompose}) for ${\mathbf J}$ and ${\mathbf B}$. We note that Eq~(\ref{eqsinchi}) is independent of a specific model for the current density ${\mathbf J}$ and the magnetic field ${\mathbf B}$ inside the flux rope. It holds for any flux rope structure (fig \ref{flrp}), and does not make any assumptions about whether or not it is force-free. We use two different methods to calculate the local pitch angle for the flux rope magnetic field, which we describe herewith.

\subsubsection{Method 1}

The force-free Lundquist solution \citep{lundquist1950} is by far the most popular concept for describing the structure of flux ropes. A natural starting point would be to assume that Lorentz self-forces arise from a situation where the flux rope structure deviates very little from the force-free state. In Eq (51) of their paper, \citet{kmrst96} give the following expression for $\sin \chi$:
\begin{equation}
 \sin \chi = \frac{\pi}{x_0} \left(\frac{a}{\pi R} \right) = \frac{\kappa}{x_0} \, ,
\label{eqrksinchi}
\end{equation}
where $x_{0} = 2.405$ is the first zero of the Bessel function $J_{0}$ and we have used Eq~(\ref{eqkappa}). Apart from the assumptions regarding conservation of axial magnetic flux and helicity, this expression for $\sin \chi$ from \citet{kmrst96} relies crucially on the assumption that the flux rope deviates very little from the force-free Lundquist solution \citep{lundquist1950}. It can be derived from Eqs~(16) and (50) of \citet{kmrst96} and using $ | {\mathbf J} | \,  | {\mathbf B} | = \alpha B^{2}$, which expresses the fact that the flux rope is nearly force-free.

Equating (\ref{eqsinchi}) and (\ref{eqrksinchi}) gives

\begin{equation}
\frac{1 - \frac{c_{2}}{c_{1}}\, \tan ^2 \gamma}{\biggl [(1 + \tan ^2 \gamma) \, (1+ \frac{c_2^2}{c_1^2} \tan ^2 \gamma) \biggr ]^{1/2}} =  \frac{\kappa}{x_0} \, .
\label{eqfirst}
\end{equation}

For the Lundquist force-free solution (e.g., Eq 1, \citealp{kmrst96}; Eq 19, \citealp{linetal98}), since ${\mathbf J} = \alpha \, {\mathbf B}$, the ratio $c_{2}/c_{1}$ (Eq~\ref{eqc1c2def}) is given by
\begin{equation}
\frac{c_{2}}{c_{1}} \equiv \frac{J_{t}}{B_{p}} \frac{B_{t}}{J_{p}} = \biggl ( \frac{J_{0}(x_{0} y)}{J_{1}(x_{0} y)} \biggr )^{2} \, 
\label{eqc2c1forcefree}
\end{equation}
where $y$ is the fractional minor radius of the flux rope. In other words, $y < 1$ defines the interior of the flux rope and $y > 1$ its exterior. The quantity $J_{0}$ denotes the Bessel function of zeroth order while the quantity $J_{1}$ represents the Bessel function of first order.
Using equation~(\ref{eqc2c1forcefree}) for $c_{2}/c_{1}$ and the observationally determined values of the similarity parameter $\kappa$ (Table \ref{tbl1}), we can use Eq~(\ref{eqfirst}) to determine the pitch angle $\gamma$ of the magnetic field configuration of a flux rope which deviates only slightly from a force-free configuration.

On the other hand, the pitch angle $\gamma$ for a the ideal force-free Lundquist solution is given by (e.g., Eq 1, \citealp{kmrst96}; Eq 19, \citealp{linetal98})

\begin{equation}
\tan \gamma  \equiv \frac{B_{p}}{B_{t}} = \frac{J_1(x_0 \, y)}{J_0(x_0 \, y)}\, ,
\label{eqforcefree}
\end{equation}

Figure \ref{Ga} shows the pitch angle calculated using equations~(\ref{eqfirst}) and (\ref{eqc2c1forcefree}), with some of the observed values of the similarity parameter $\kappa$ for all the CMEs in our list (table \ref{tbl1}).  The blue line in figure \ref{Ga} uses the smallest value of $\kappa$ observed in our sample ($\kappa = 0.2$), while the green line uses the largest observed value of $\kappa$ (= 0.44). For comparison, the pitch angle computed using the ideal force-free configuration (Eq \ref{eqforcefree}) is also overplotted in red. Clearly, the local magnetic field pitch angles for self-similarly expanding flux ropes do not agree with that for an ideal force-free configuration; the larger the value of $\kappa$, the more the disagreement. In other words, the magnetic field configurations in the observed (self-similarly expanding) flux ropes deviate considerably from a force-free one. This is despite the fact that the observed values of the self-similarity parameter ($\kappa$, Table 1) correspond to misalignment angles $\chi$ (equation~\ref{eqrksinchi}) of only 5$^{\circ}$ to 10$^{\circ}$. The nearly force-free assumption is thus not consistent, and it is worth examining if such self-similarly expanding flux ropes can be better described by a non-force free model.

\subsubsection{Method 2}
We consider a prescription for a non-force-free flux rope configuration given by \citet{ber13}. This prescription is a perturbative expansion on a force-free configuration, correct to order $\kappa \equiv a/R$, which incorporates the effect of large-scale curvature (see Fig \ref{flrp}). In this prescription, the pitch angle is defined as

\begin{equation}
\tan \gamma \, \equiv \, \frac{B_p}{B_t} \, = \, \frac{J_1\left( A(y, \phi)\right)}{J_0\left( A(y, \phi)\right)} \, ,
\label{eqberdi1}
\end{equation}
where $\phi$ is the polar angle coordinate in the plane perpendicular to the toroidal axis and the quantity $A$ is defined by

\begin{equation}
A (y, \phi) = x_{o}\, y \left[  1+ \kappa y \, (\cos \phi - | \sin \phi |) \right ] \, .
\label{eqberdi2}
\end{equation}

The quantity $A (y, \phi)$ expresses the effect of the curvature of the major radius. For a straight flux rope, an observer looking through the cross section will see only one circle, because the circles defining the cross section overlap each other. For a bent flux rope, on the other hand, the observer will see a few circles displaced from each other. The more the flux rope curvature, the farther the centers of these circles are displaced from each other. Upon comparing equations~(\ref{eqberdi1}) and (\ref{eqforcefree}), it is evident that the non-force free expression is identical to the force-free expression for $\phi = \pi/4$ and $3 \pi/4$. Using Eqs~(\ref{eqberdi1}) and (\ref{eqberdi2}), we can calculate the local magnetic field pitch angle ($ \gamma$) for this non-force-free configuration as a function of the observed similarity parameter $\kappa$ (table \ref{tbl1}). The local magnetic field pitch angle $\gamma$ is depicted as a function of the fractional minor radius $y$ for a few representative values of $\kappa$ and $\phi$ in figure \ref{TGXk}. The red solid line is for $\phi=\pi/4$, the blue dotted line is for $\phi=\pi/2$ and $\kappa=0.2$ while the blue solid line is for $\phi=\pi/2$ and $\kappa=0.44$.  Since $\phi = \pi/4$ corresponds to the force-free case, it is independent of $\kappa$. Using the values of $\gamma$ shown in figure \ref{TGXk}, we can compute the misalignment angle $\chi$ between ${\mathbf J}$ and ${\mathbf B}$ using equation~(\ref{eqsinchi}). However, since we are considering a non-force free configuration, it is not appropriate to use equation~(\ref{eqc2c1forcefree}) for the ratio $c_{2}/c_{1}$, which is valid only for a force-free configuration. Instead, we recognize that the poloidal magnetic field is generated by a toroidal current $I_{t}$, while the toroidal magnetic field is generated by a poloidal current $I_{p}$:

\begin{equation}
B_p = \frac{2I_t}{ca} \, , \,\,\,\,\,B_t = \frac{2I_p}{cR} \, ,
\label{eqJB}
\end{equation}

where $c$ denotes the speed of light. Furthermore, the total currents are related to their respective densities by (e.g., \citealp{chn89})

\begin{equation}
I_t  = 2 \pi \int_0^a r J_t dr = \pi a^2 J_t  \, , \,\,\,\,\, I_p = 2 \pi R \int_0^a J_p dr = 2 \pi R a J_p \, ,
\label{eqIJ}
\end{equation}
where, for the sake of concreteness, we have assumed that the current density is uniform throughout the body of the flux rope. We note that other current distributions are possible. Equations (\ref{eqc1c2def}), (\ref{eqJB}) and (\ref{eqIJ}) yield

\begin{equation}
\frac{c_{2}}{c_{1}} = 2 \, .
\label{eqc1c2}
\end{equation}

Using the values of $\tan \gamma$ from equations (\ref{eqberdi1}) and (\ref{eqberdi2}) in equation~(\ref{eqsinchi}) with the quantity $c_{2}/c_{1}$ given by equation~(\ref{eqc1c2}), we get the values of $\chi$ depicted in figure~\ref{CX} for the same parameter values used in figure \ref{TGXk}. The linestyles are identical to those used in figure \ref{TGXk}. Clearly, the angle between ${\mathbf J}$ and ${\mathbf B}$ can be substantial, which means that the flux rope deviates considerably from a force-free state.

\section{Summary}
It is generally accepted that Lorentz self-forces are responsible for the evolution of CMEs (both expansion as well as translation). While Lorentz self-forces are assumed to arise from misaligned current density (${\mathbf J}$) and magnetic field (${\mathbf B}$), the degree of misalignment is not yet clear. In order to remedy this, we have derived a general relation (Eq~\ref{eqsinchi}) between the local pitch angle ($\gamma$) of the flux rope magnetic field and the misalignment angle ($\chi$) between ${\mathbf J}$ and ${\mathbf B}$.

We have fitted the \citet{thr09} 3D flux rope model to nine well observed CMEs in the SECCHI/STEREO field of view. One of the main conclusions from this exercise is that the flux rope CMEs propagate in a nearly self-similiar manner; i.e., the ratio ($\kappa$) of the flux rope minor to major radius remains approximately constant as it propagates outwards. This conclusion is consistent with those from similar exercises using COR2 data (\citealp{kilpua2012}) and HI data (\citealp{robin2013}).

We have used the observed values of the self-similarity parameter ($\kappa$) to calculate the local pitch angle of the flux rope magnetic field ($\gamma$) using two different prescriptions. 
In the first one, we have assumed that the flux rope deviates only slightly from a force-free equilibrium following the prescription of \citet{kmrst96}. Even though this prescription predicts that the misalignment angle ($\chi$) between ${\mathbf J}$ and ${\mathbf B}$ is only 5$^{\circ}$ to 10$^{\circ}$, the local magnetic field pitch angle calculated deviates appreciably from that calculated using the purely force-free assumption (figure \ref{Ga}). This implies that the nearly force-free assumption is not well justified.

We therefore adopt a second method that employs an explicit expression for magnetic fields in a non-force-free flux rope configuration. This is a first order perturbation (in the quantity $\kappa$) to a force-free flux rope \citep{ber13}. This method yields values of the local magnetic field pitch angle as a function of radial position inside the flux rope as well as the azimuthal angle (figure \ref{TGXk}). Since the second method does not assume a priori that the flux rope is nearly force-free, we contend that results using this method (figures \ref{TGXk} and \ref{CX}) are more reliable. 
The values for the angle ($\chi$) between ${\mathbf J}$ and ${\mathbf B}$ deduced from method 2 (figure \ref{CX}) are substantial; they range from $-90^{\circ}$ to $90^{\circ}$. These values may be contrasted with the rather small values for $\chi$ (around $3^{\circ}$) that are required for flux rope prominences to be supported against gravity (\citealp{rustkumar94}). 


\section{Conclusions}
Our findings imply that, in the coronagraph field of view, the current (${\mathbf J}$) and the magnetic field (${\mathbf B}$) within flux rope CMEs that propagate in a self-similar manner can be substantially misaligned. This is the first conclusive evidence of the non force-free nature of flux rope CMEs, which forms the basis for Lorentz self-force driving. The magnitude of the Lorentz self force ($|{\mathbf J}|\,|{\mathbf B}|\,\sin \chi$) depends upon the magnitudes of the current density ($|{\mathbf J}|$) and magnetic field ($|{\mathbf B}|$) as well as the the angle ($\chi$) they subtend to each other. The results from this work regarding $\chi$ can constrain the magnitudes of the current and magnetic field in flux ropes needed to explain an observationally mandated driving force (e.g., \citealp{sv07}). Furthermore, our findings imply that there is excess magnetic energy that is available within the flux rope structure which can be expended in translating and expanding the CME, in (often) driving a shock ahead of it, and in heating the plasma that is inside it. It is not yet clear how the available magnetic energy is partitioned among these different avenues.

\begin{figure}
\includegraphics[width = \columnwidth]{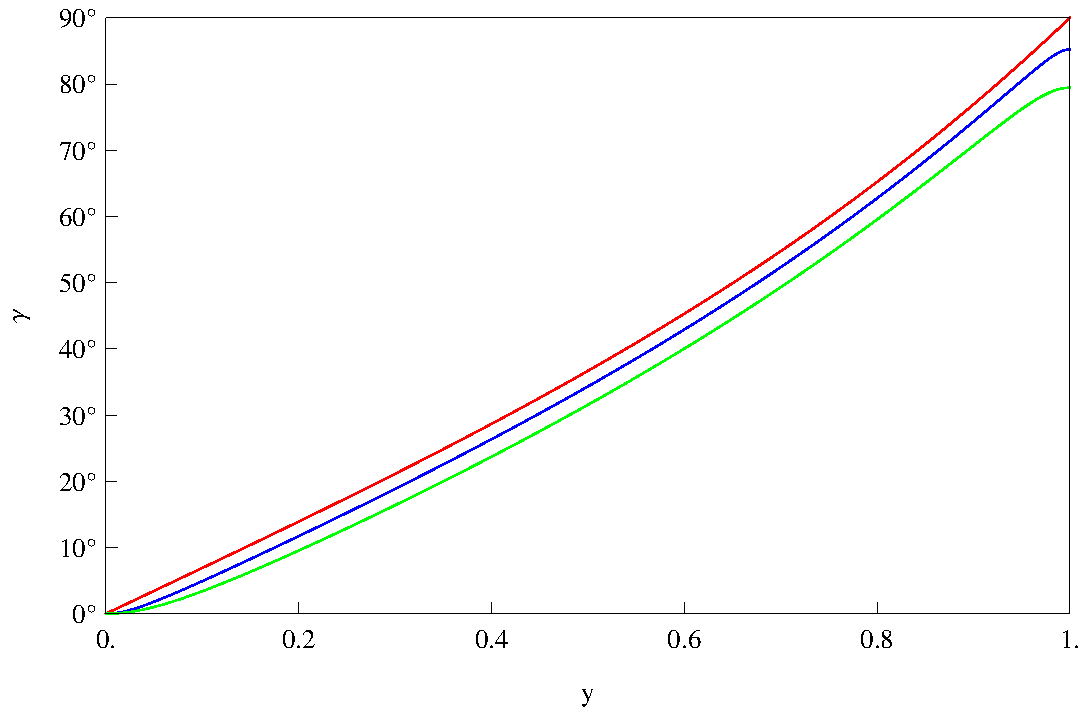}
\caption{A plot of the local magnetic field pitch angle $\gamma$ as a function of fractional minor radius $y$. The red line denotes the force free model (Eq~\ref{eqforcefree}) while the blue and green lines are obtained using method 1 (Eq~\ref{eqfirst}). The blue line uses $\kappa = 0.2$ and the green line employs $\kappa = 0.44$} 
\label{Ga}
\end{figure}

\begin{figure}
\includegraphics[width = \columnwidth]{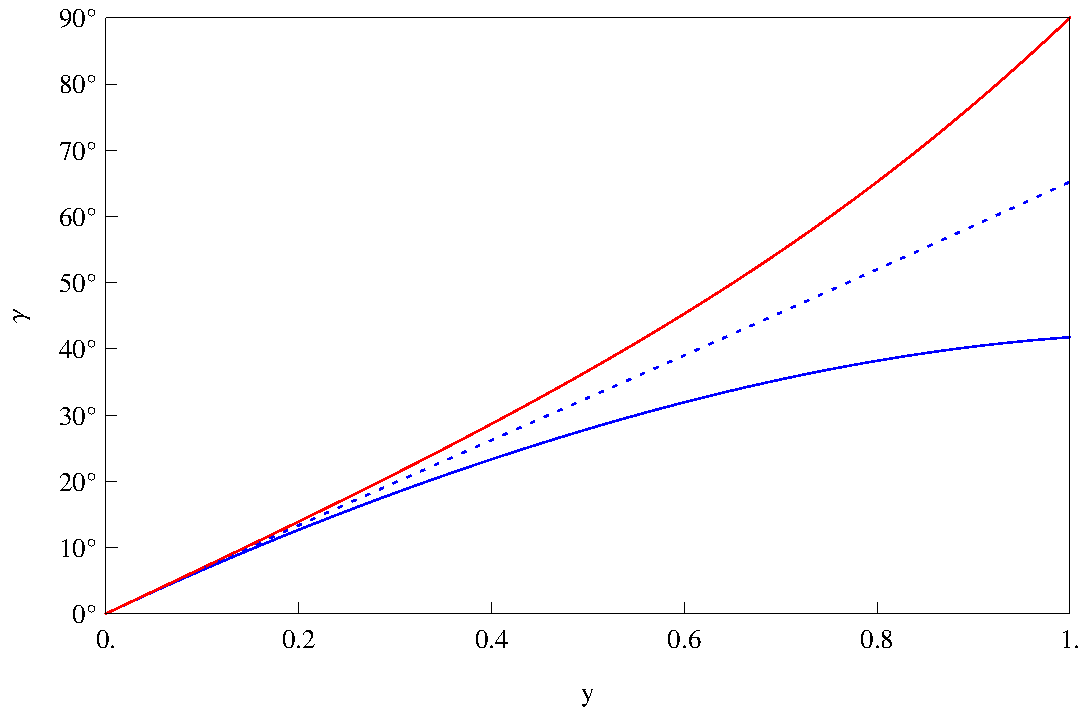}
\caption{A plot of $\gamma$ as a function of fractional minor radius $y$  using method 2 (Eq~\ref{eqberdi1}). The red solid line is for $\phi=\pi/4$, the blue dotted line is for $\phi=\pi/2$ and $\kappa=0.2$ while the blue solid line is for $\phi=\pi/2$ and $\kappa=0.44$.}
\label{TGXk}
\end{figure}

\begin{figure}
\includegraphics[width = \columnwidth]{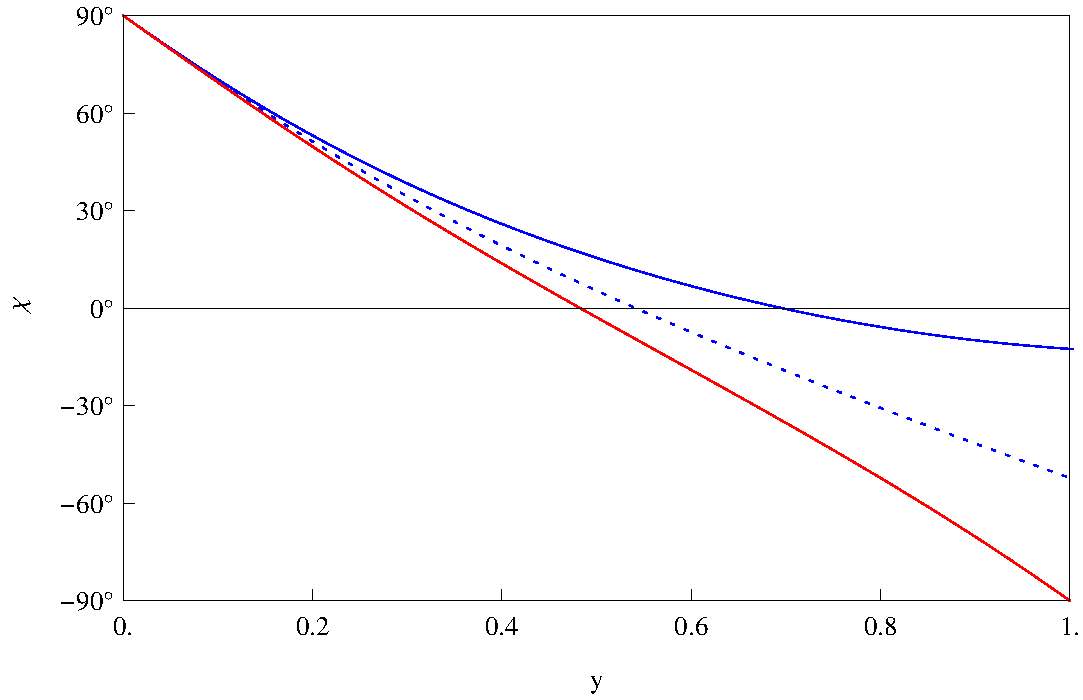}
\caption{A plot of the angle ($\chi$) between ${\mathbf J}$ and ${\mathbf B}$ as a function of fractional minor radius $y$ using method 2. The linestyles are the same as that used in figure~\ref{TGXk}.}
\label{CX}
\end{figure}

\begin{figure}
\includegraphics[width = \columnwidth]{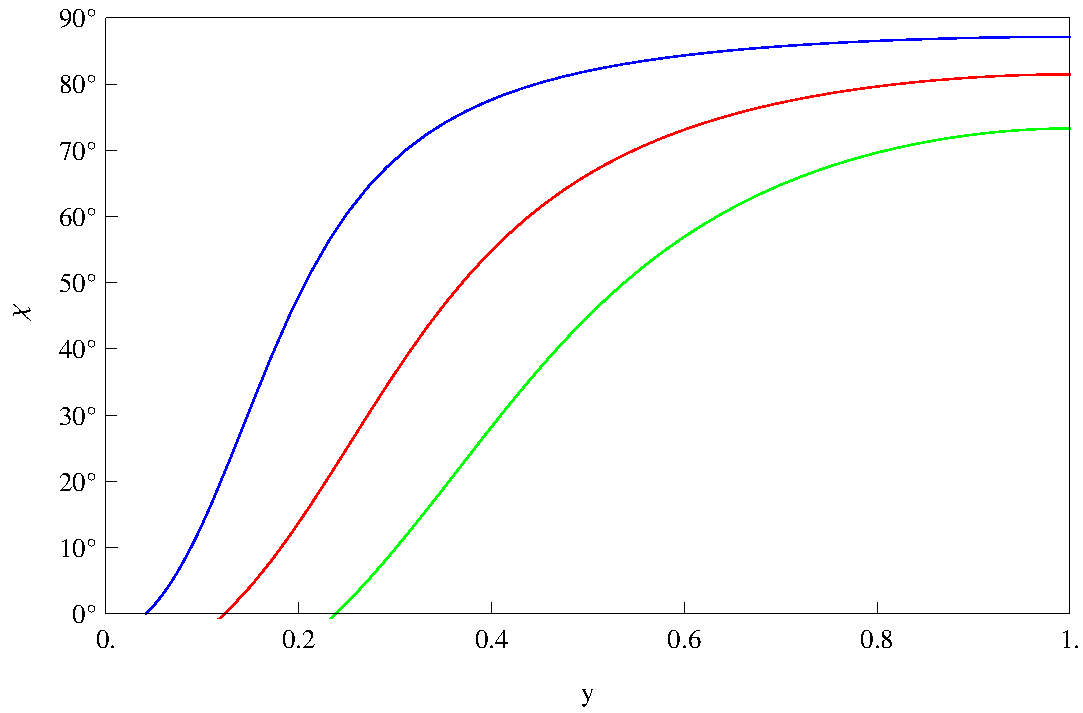}
\caption{The misalignment angle ($\chi$) between the current density (${\mathbf J}$) and the magnetic field (${\mathbf B}$) implied by inferences of the magnetic field pitch angle ($\gamma$) near the Earth. The misalignment angle is plotted as a function of the fraction radius ($y$) inside the flux rope. The blue curve is plotted for $\tan \gamma = 0.05$, the red one for $\tan \gamma = 0.15$ and the green one for $\tan \gamma = 0.3$}
\label{NearEarth}
\end{figure}

Finally, it is worth comparing our estimates for the magnetic field pitch angle (figures \ref{Ga} and \ref{TGXk}) with values for this quantity near the Earth. 
Observations of near-Earth magnetic clouds (which are generally modeled as force-free flux ropes) suggest that $0.05 \lesssim \tan \gamma \lesssim 0.3$ \citep{lar91, lea04, guli05}. Along with plausible guesses for the the number of field line turns, these values for $\tan \gamma$ have been used to infer total field line lengths in near-Earth magnetic clouds \citep{kah11}, which in turn are  used to address questions related to whether or not a force-free flux rope configuration is a good model for these structures, and if their legs are still connected to the Sun. In keeping with the general expectation that flux ropes observed near the Earth are force-free structures, we use method 1 (which relies on the nearly force-free assumption) to check if these observed magnetic field pitch angles are consistent with small values of the misalignment angle $\chi$, as they are often assumed to be. We use equations~(\ref{eqsinchi}) and (\ref{eqc2c1forcefree}) to calculate the values of $\chi$ implied by the observed range $0.05 \lesssim \tan \gamma \lesssim 0.3$. The results are shown in figure~(\ref{NearEarth}). This figure shows the misalignment angle ($\chi$) between the the current density and the magnetic field inside the flux rope corresponding to $\tan \gamma = 0.05$ (blue line), $\tan \gamma = 0.15$ (red line) and $\tan \gamma = 0.3 $ (green line). This shows that, even near the Earth, the angle ($\chi$) between ${\mathbf J}$ and ${\mathbf B}$ is often quite substantial, and the force-free assumption is probably not valid.


\acknowledgments 

We acknowledge insightful comments from the anonymous referee that have helped improve the paper. PS and KPA acknowledge support from the Asian Office of Aerospace
Research and Development, Tokyo. We thank Bhavesh Khamesra and Nishtha Sachdeva for
help with analysing the SECCHI data.
AV is supported by NASA contract
S-136361-Y to NRL.  STEREO is the third mission in NASA's Solar
Terrestrial Probes program. The SECCHI data are produced by an
international consortium of the NRL, LMSAL and NASA GSFC (USA), RAL
and University of Birmingham (UK), MPS(Germany), CSL (Belgium), IOTA
and IAS (France).






\clearpage



\begin{deluxetable}{llcccccc}
\tablewidth{0pt}
\tablecaption{Flux rope fits to STEREO COR2 data\label{tbl1}}
\tablehead{
\colhead{Date}           & \colhead{Time}      &
\colhead{Longitude}          & \colhead{Latitude}  &
\colhead{Tilt Angle}          & \colhead{Height}    &
\colhead{$\kappa \, = \, \frac{a}{R}$}  } 
\startdata
07/01/2010 & 07:07 & 135.28 & 10.06 & 3.35 & 7.50 &   \\
07/01/2010 & 08:08 & 134.16 & 09.50 & 3.35 & 8.43 &   \\
07/01/2010 & 09:08 & 133.05 & 09.50 & 3.91 & 9.50 &     \\
07/01/2010 & 10:08 & 131.93 & 10.62 & 3.35 & 10.21 &   {0.212$\pm$0.008}\\
07/01/2010 & 11:08 & 133.05 & 10.06 & 0.56 & 10.86 & \\
07/01/2010 & 12:08 & 133.05 & 10.62 & 2.80 & 12.14 &  \\ \hline \hline
01/02/2010 & 19:08 & 38.01 & -19.01 & 14.54 & 10.21 &  \\
01/02/2010 & 20:08 & 36.90 & -19.57 & 16.77 & 12.21 &  {0.315$\pm$0.006}\\
01/02/2010 & 21:08 & 35.78 & -19.01 & 17.89 & 14.57 & \\
01/02/2010 & 22:08 & 35.78 & -19.01 & 19.01 & 16.5 &  \\ \hline \hline
14/02/2010 & 03:08 & 210.18 & 13.98 & -34.66 & 8.14 &  \\
14/02/2010 & 04:08 & 210.18 & 12.30 & -30.75 & 9.79 &  {0.258$\pm$0.010}\\
14/02/2010 & 05:08 & 210.18 & 11.74 & -40.25 & 11.14 &  \\
14/02/2010 & 06:08 & 211.31 & 11.74 & -38.57 & 13.00 &  \\ \hline \hline
12/06/2010 & 15:08 & 338.76 & 35.22 & 72.67 & 9.21 &  \\
12/06/2010 & 16:08 & 336.52 & 34.10 & 72.67 & 11.29 &   {0.305$\pm$0.024}\\
12/06/2010 & 17:08 & 336.52 & 29.07 & 77.14 & 13.00 &   \\
12/06/2010 & 18:08 & 336.52 & 29.07 & 77.14 & 15.71 &   \\ \hline \hline
20/06/2010 & 22:08 & 310.81 & 11.18 & 0.56 & 7.14 &   \\
20/06/2010 & 23:08 & 310.81 & 11.18 & 0.56 & 8.14 &  \\
21/06/2010 & 00:08 & 310.81 & 11.18 & 0.56 & 9.07 &  {0.196$\pm$0.013}\\
21/06/2010 & 01:08 & 310.81 & 11.18 & 0.56 & 10.43 &   \\
21/06/2010 & 02:08 & 310.81 & 12.30 & 0.56 & 11.64 &   \\ \hline \hline
01/03/2010 & 05:08 & 24.60 & -16.77 & 3.35 & 10.57 &  \\
01/03/2010 & 06:08 & 24.60 & -16.77 & 3.35 & 12.50 &  {0.353$\pm$0.010}  \\
01/03/2010 & 07:08 & 24.60 & -16.77 & 2.80 & 13.93 &   \\
01/03/2010 & 08:08 & 23.48 & -15.65 & -3.91 & 15.93 &   \\ \hline \hline
26/03/2010 & 13:08 & 22.36 & -1.12 & 46.96 & 8.79 &   \\
26/03/2010 & 14:08 & 22.36 & -1.12 & 51.99 & 10.14 &   \\
26/03/2010 & 15:08 & 22.36 & -1.12 & 54.78 & 11.43 &   {0.216$\pm$0.011}\\
26/03/2010 & 16:08 & 22.36 & -0.56 & 55.90 & 13.21 &   \\
26/03/2010 & 17:08 & 22.36 & -1.12 & 87.20 & 14.71 &   \\ \hline \hline
13/04/2010 & 12:08 & 164.34 & 36.33 & -12.30 & 6.50 &  \\
13/04/2010 & 13:08 & 164.34 & 34.66 & -11.18 & 9.07 &  {0.438$\pm$0.032} \\
13/04/2010 & 14:08 & 164.34 & 34.10 & -14.54 & 12.07 &   \\
13/04/2010 & 15:08 & 164.34 & 33.54 & -13.98 & 15.98 &   \\ \hline \hline
29/01/2008 & 06:22 & 54.78 & 3.91 & -0.56 & 11.21 &   \\
29/01/2008 & 06:52 & 55.90 & 3.91 & -0.56 & 11.71 &   \\
29/01/2008 & 07:22 & 55.90 & 3.91 & -0.56 & 12.35 &  {0.203$\pm$0.008}\\
29/01/2008 & 07:52 & 55.90 & 3.35 & -0.56 & 13.29 &   \\
29/01/2008 & 08:22 & 55.90 & 3.91 & -0.56 & 13.86 &   \\
29/01/2008 & 09:22 & 55.90 & 4.47 & -1.12 & 15.71 &  \\ \enddata
\end{deluxetable}


\begin{thebibliography}{}  
\bibitem[Anzer \& Poland(1979)]{anz} Anzer, U., Poland, A. I. 1979, Solar Phys., 61, 95
\bibitem[Berdichevsky(2013)]{ber13} Berdichevsky, D. B.  2013, Solar Phys., 284, 245
\bibitem[Cargill(2004)]{cgl04} Cargill, P. J. 2004, Solar Phys., 221, 135
\bibitem[Chen(1989)]{chn89} Chen, J. 1989, \apj, 344, 1051
\bibitem[Chen(1996)]{chn96} Chen, J. 1996, \jgr 101, 27499
\bibitem[Chen(2012)]{chn12} Chen, J. 2012, \apj, 761, 179
\bibitem[Chen \& Krall(2003)]{chkr} Chen, J., Krall, J. 2003, \jgr, A108, 1410
\bibitem[Cheng et al(2014)]{cheng2014} Cheng, X., Ding, M. D., Guo, Y., Zhang, J., Vourlidas, A., Liu, Y. D., Olmedo, O., Sun, J. Q., Li, C. 2014, \apj, 780, 28
\bibitem[Colaninno, Vourlidas \& Wu(2013)]{robin2013} Colaninno, R. C., Vourlidas, A., Wu, C.-C 2013, \jgr, 118, 6866
\bibitem[Demoulin \& Dasso(2009)]{Dem} D\'emoulin, P., Dasso, S. 2009, \aa, 498, 551
\bibitem[Garren \& Chen(1994)]{garchen} Garren, D. A., Chen, J. 1994, PhPl, 1, 3425
\bibitem[Gopalswamy et al.(2000)]{gpl00} Gopalswamy, N., Lara, A., Lepping, R. P., Kaiser, M. L., Berdichevsky, D., St. Cyr, O. C. 2000, GeoRL, 27, 145
\bibitem[Gulisano et al.(2005)]{guli05} Gulisano, A. M., Dasso, S., Mandrini, C. H., Demoulin, P. 2005, J. Atmos. Terr. Phys., 67, 1761

\bibitem[Howard et al.(2008)]{howard08} Howard, R. A. et al. 2008, Sp. Sci. Rev., 136, 67

\bibitem[Kaiser et al.(2008)]{kaiser08} Kaiser, M. L. et al 2008, Sp. Sci. Rev., 136, 5 
\bibitem[Kahler, Haggerty \& Richardson (2011)]{kah11} Kahler, S. W., Haggerty, D. K., Richardson, I. G. 2011, \apj, 106, doi:10.1088/0004-637X/736/2/106

\bibitem[Kilpua et al. (2012)]{kilpua2012} Kilpua, E. K. J., Mierla, M., Rodriguez, L., Zhukov, A. N., Srivastava, N., West, M. J. 2012, Solar Phys., 279, 477

\bibitem[Krall, Chen, \& Santoro (2000)]{kral00} Krall, J., Chen, J., Santoro, R. 2000, \apj, 539, 964
\bibitem[Krall et al.(2001)]{kral01} Krall, J., Chen, J., Duffin, R. T., Howard, R. A., Thompson, B. J. 2001, \apj, 562, 1045
\bibitem[Kumar \& Rust(1996)]{kmrst96} Kumar, A., Rust, D. M., 1996, \jgr 101, 15667
\bibitem[Landau \& Lifshitz(1984)]{lanlif} Landau L.D., Lifshitz E.M.  \& Pitaevskii L. P., Electrodynamics of Continuous Media, 2nd ed., Pergamon, Tarrytown, N.Y., 1984.
\bibitem[Larson et al.(1997)]{lar91} Larson, D. E., et al 1997, \grl, 24, 1911
\bibitem[Leamon et al.(2004)]{lea04} Leamon, R. J., Canfield, R. C.; Jones, S. L.; Lambkin, K.; Lundberg, B. J.; Pevtsov, A. A., 2004, \jgr A109, 5106
\bibitem[Lewis \& Simnett(2002)]{lew02} Lewis, D. J., Simnett, G. M., 2002, MNRAS, 333, 969
\bibitem[Lin et al (1998)]{linetal98} Lin, J., Forbes, T. G., Isenberg, P. A. \& D\'emoulin, P. 1998, \apj, 504, 1006
\bibitem[Lundquist(1950)]{lundquist1950} Lundquist, S., 1950, Ark. Fys., 2, 361
\bibitem[Olmedo \& Zhang (2010)]{olmedo2010} Olmedo, O., Zhang, J. 2010, \apj, 718, 433
\bibitem[Olmedo et al.(2013)]{olmetal} Olmedo, Oscar, Zhang, Jie, Kunkel, Valbona 2013, \apj, 771, 125O
\bibitem[Poomvises et al.(2010)]{poom10} Poomvises, W., Zhang, J., Olmedo, O. 2010, \apjl, 717, L159
\bibitem[Rust \& Kumar(1994)]{rustkumar94} Rust, D. M., Kumar, A. 1994, Solar Phys., 155, 69
\bibitem[Shafranov(1966)]{sha66} Shafranov V. D., 1966, Reviews of Plasma Physics, 2, 103
\bibitem[Song et al.(2013)]{sng13} Song, H. Q., Chen, Y., Ye, D. D., Han, G. Q., Du, G. H., Li, G., Zhang, J., Hu, Q. 2013, \apj, 773, 129
\bibitem[Subramanian, Lara \& Borgazzi(2012)]{slb12} Subramanian, P.,Lara, A., Borgazzi 2012, \grl,  39, L19107
\bibitem[Subramanian \& Vourlidas(2007)]{sv07} Subramanian, P., Vourlidas, A. 2007, \aap, 467, 685
\bibitem[Subramanian \& Vourlidas(2009)]{sv09} Subramanian, P., Vourlidas, A. 2009, \apj, 693, 1219
\bibitem[Thernisien$,$ Vourlidas \& Howard(2009)]{thr09} Thernisien, A., Vourlidas, A., Howard, R. A., 2009, Solar Phys. 256, 111T
\bibitem[Vourlidas et al(2011)]{angelos2011} Vourlidas, A., Colaninno, R.,  Nieves-Chinchilla, T., Stenborg, G. 2011, \apjl, 733, L23
\bibitem[Vourlidas et al.(2000)]{vr00} Vourlidas, A., Subramanian, P., Dere, K. P., Howard, R. A., 2000, \apj, 534, 456V
\bibitem[Vourlidas et al.(2013)]{angelos13} Vourlidas, A., Lynch, B. J., Howard, R. A., Li, Y. 2013, Solar Phys., 284, 179
\bibitem[Vrsnak(2006)]{vrs} Vrsnak, B. 2006, AdSpR, 38, 431
\bibitem[Wang et al.(2009)]{wang09} Wang, Y., Zhang, J., Shen, C. 2009, \jgr, 114, A10104
\bibitem[Webb \& Howard(2012)]{webb12} Webb, D. F. and Howard, T. A. 2012, Living Rev. in Solar Phys., 9, 3
\bibitem[Yashiro et al.(2004)]{yashiro04} Yashiro, S. et al 2004, \jgr, 109, 07105
\bibitem[Zhang et al.(2013)]{jie13} Zhang, J., Hess, P., Poomvises, W. 2013, Solar Phys., 284, 89

\end{thebibliography}
\end{document}